\catcode`\@=11

\expandafter\ifx\csname draftcontrol\endcsname\relax\global\def\draftcontrol{0}
\fi

{\count255=\time\divide\count255 by 60
\xdef\hourmin{\number\count255}
\multiply\count255 by-60\advance\count255 by\time
\xdef\hourmin{\hourmin:\ifnum\count255<10 0\fi\the\count255}}
\def\draftdate{\number\month/\number\day/\number\year\ \ \ \hourmin }

\newcommand\makepapertitle{\par
  \begingroup
    \renewcommand\thefootnote{\@fnsymbol\c@footnote}%
    \def\@makefnmark{\rlap{\@textsuperscript{\normalfont\@thefnmark}}}%
    \long\def\@makefntext##1{\parindent 1em\noindent
            \hb@xt@1.8em{%
                \hss\@textsuperscript{\normalfont\@thefnmark}}##1}%
     \newpage
     \global\@topnum\z@   
     \@makepapertitle
     \thispagestyle{empty}\@thanks
  \endgroup
  \setcounter{footnote}{0}%
  \global\let\thanks\relax
  \global\let\makepapertitle\relax
  \global\let\@makepapertitle\relax
  \global\let\@thanks\@empty
  \global\let\@author\@empty
  \global\let\@date\@empty
  \global\let\@title\@empty
  \global\let\title\relax
  \global\let\author\relax
  \global\let\date\relax
  \global\let\and\relax
  \def\version{\let\version\@version\@gobble}
}
\def\@makepapertitle{%
  \newpage
   \ifnum\draftcontrol=1 {}
   \version\versionno
   \vskip 3em%
   \else
   \hfill\hbox to 3cm {\parbox{4cm}{\@pubnum}\hss}%
   \vskip 3em%
   \fi
   \begin{center}%
   \let \footnote \thanks
     {\LARGE \@title \par}%
     \vskip 1.5em%
     {\normalsize
       \lineskip .5em%
       \begin{center} 
         \@author
       \end{center} 
\par}%
     \vskip 1em%
     {\@bstract}%
     \end{center}%
     \vskip .5em
     \@date%
   \par
}

\gdef\@pubnum{}
\def\pubnum#1{%
  \gdef\@pubnum{#1}}

\gdef\@bstract{}
\def\Abstract#1{%
  \gdef\@bstract{%
   \parbox{\textwidth-0pc}{%
   \centerline{\bf Abstract}\penalty1000
   \noindent
   \renewcommand\baselinestretch{1.0}
   {#1}}}
}

\def\ps@paper{\let\@mkboth\@gobbletwo%
     \ifnum\draftcontrol=1
        \def\@oddfoot{\hbox to \textwidth{\tiny \versionno \hfil\tiny\draftdate}%
        \hskip -\textwidth \hbox to \textwidth{\hfil\rm\thepage\hfil}}%
     \else\def\@oddfoot{\hbox to \textwidth{\hfil\rm\thepage\hfil}}
     \fi
     \let\@evenfoot\@oddfoot
}

\def\body{\clearpage
          \pagestyle{paper}
        }
\newenvironment{acknowledgments}{%
\vskip 3.25ex
\noindent {\bf Acknowledgments}
}

\def\@version#1{\ifnum\draftcontrol=1
\typeout{}\typeout{#1}\typeout{}
\vskip3mm\centerline{\hbox{\fbox{\normalsize{\tt DRAFT -- #1 -- }
                   {\draftdate}}}}\vskip3mm
\fi}
\let\version\@version
\long\def\eqlabel#1{\ifnum\draftcontrol=1
                    \tag@false  
                    \tag*{(\theequation) \hbox to -0.2cm{\hspace{0cm}\small{#1}\hss}}
                    \refstepcounter{equation} 
                    \edef\@currentlabel{\theequation}
                    \ltx@label{#1}          
                    \else
                    \label{#1}
                    \fi
                    }
\let\st@bibitem\@bibitem
\let\st@lbibitem\@lbibitem
\ifnum\draftcontrol=1
  \def\@bibitem#1{%
    \st@bibitem{#1}\a@@label{#1}\ignorespaces}
  \def\@lbibitem[#1]#2{%
    \st@lbibitem[#1]{#2}\a@@label{#2}\ignorespaces}
  \def\a@@label#1{%
    \gdef\a@lab{\smash{\normalfont\small#1}}
    \ifvmode
      \if@inlabel
        \global\setbox\@labels\hbox{%
          \llap{\a@lab\let\a@lab\relax
                \kern\@totalleftmargin\kern\marginparsep}%
          \box\@labels}%
      \fi
    \fi}
\fi

\documentclass[12pt,letterpaper]{article}

\usepackage{amsmath,amssymb,array,calc,rotating,epsfig,psfrag}
\usepackage[nosort]{cite}

\ifnum\draftcontrol=1
\fi

\renewcommand\baselinestretch{1.25}
\renewcommand\section{\@startsection {section}{1}{\z@}%
                                   {-3.5ex \@plus -1ex \@minus -.2ex}%
                                   {2.3ex \@plus.2ex}%
                                   {\normalfont\large\bfseries}}
\renewcommand\subsection{\@startsection{subsection}{2}{\z@}%
                                     {-3.25ex\@plus -1ex \@minus -.2ex}%
                                     {1.5ex \@plus .2ex}%
                                     {\normalfont\normalsize\bfseries}}
\renewcommand\subsubsection{\@startsection{subsubsection}{3}{\z@}%
                                     {-3.25ex\@plus -1ex \@minus -.2ex}%
                                     {1.5ex \@plus .2ex}%
                                     {\normalfont\normalsize\it}}

\numberwithin{equation}{section}

\def\projective   {{\mathbb P}}

\def\reals        {{\mathbb R}}

\def\del          {\partial}

\def\revise#1       {\marginpar{\rule{2mm}{1cm} #1}}

\def\RR{\reals}
\def\PP{\projective}
\def\RP{\RR\PP}

\def\R{{\rm R}}

\def\sqr#1#2{{\vcenter{\vbox{\hrule height.#2pt  
 \hbox{\vrule width.#2pt height#1pt \kern#1pt
 \vrule width.#2pt}\hrule height.#2pt}}}}

\def\yboxit#1#2{\vbox{\hrule height #1 \hbox{\vrule width #1
\vbox{#2}\vrule width #1 }\hrule height #1 }}
\def\fillbox#1{\hbox to #1{\vbox to #1{\vfil}\hfil}}
\def\ybox{{\lower 1.3pt \yboxit{0.4pt}{\fillbox{8pt}}\hskip-0.2pt}}

\def\comments#1{}

\def\half{{\frac12}}
\def\pder#1#2{{\frac{\partial{#1}}{\partial{#2}}}}
\def\oder#1#2{{\frac{d{#1}}{d{#2}}}}
\def\cint#1#2{{\oint_{#1}\frac{d#2}{2\pi i}}}
\def\Tr{{{\rm Tr~ }}}
\def\tr{{\rm tr\ }}

\def\adj{{\rm adj}}

\def\odd{{\rm odd}}

\def\CF{{\cal F}}
\def\CG{{\cal G}}

\def\CM{{\cal M}}
\def\CN{{\cal N}}
\def\CO{{\cal O}}

\def\CW{{\cal W}}

\def\P{\BP}

\def\II{\relax{I\kern-.10em I}}

\def\IZ{\relax\ifmmode\mathchoice
{\hbox{\cmss Z\kern-.4em Z}}{\hbox{\cmss Z\kern-.4em Z}}
{\lower.9pt\hbox{\cmsss Z\kern-.4em Z}}
{\lower1.2pt\hbox{\cmsss Z\kern-.4em Z}}\else{\cmss Z\kern-.4em
Z}\fi}
\def\IB{\relax{\rm I\kern-.18em B}}
\def\IC{{\relax\hbox{$\inbar\kern-.3em{\rm C}$}}}
\def\ID{\relax{\rm I\kern-.18em D}}
\def\IE{\relax{\rm I\kern-.18em E}}
\def\IF{\relax{\rm I\kern-.18em F}}
\def\IG{\relax\hbox{$\inbar\kern-.3em{\rm G}$}}
\def\IGa{\relax\hbox{${\rm I}\kern-.18em\Gamma$}}
\def\IH{\relax{\rm I\kern-.18em H}}
\def\II{\relax{\rm I\kern-.18em I}}
\def\IK{\relax{\rm I\kern-.18em K}}
\def\IP{\relax{\rm I\kern-.18em P}}

\def\inbar{\,\vrule height1.5ex width.4pt depth0pt}

\font\cmss=cmss10 \font\cmsss=cmss10 at 7pt
\def\IR{\relax{\rm I\kern-.18em R}}

\def\BR{{\mathbb {R}}}
\def\BZ{{\mathbb {Z}}}
\def\BP{{\mathbb {P}}}
\def\BC{{\mathbb {C}}}

\def\lp10{l_P^{10}}
\def\lp11{l_P^{11}}

\newcommand{\nc}{\newcommand}
\nc{\rnc}{\renewcommand}
\nc{\CY}{Calabi-Yau}
\nc{\CYM}{Calabi-Yau manifold}
\nc{\CYMs}{Calabi-Yau manifolds}
\nc{\DB}{D-Brane}
\nc{\DBs}{D-Branes}
\nc{\SUSY}{supersymmetry}
\nc{\Kah}{K\"ahler}
\nc{\cs}{complex structure}
\nc{\beq}{\begin{equation}}
\nc{\eeq}{\end{equation}}
\nc{\ntwo}{${\cal N}=2$}
\nc{\nOne}{${\cal N}=1$}
\nc{\hs}{\hspace{0.2in}}
\nc{\Z}{{\mathbb Z}}
\rnc{\P}{{\mathbb P}}
\rnc{\RP}{{\mathbb {RP}}}
\nc{\WP}{\mathbb{WP}}
\nc{\slag}{special Lagrangian}
\nc{\cn}{\C^n}
\nc{\rn}{\R^n}

\nc{\SO}{SO}
\nc{\Sp}{Sp}
\nc{\SU}{SU}

\nc{\Wtree}{W_{\hbox{\small tree}}}
\nc{\Weff}{W_{\hbox{\small eff}}}

\catcode`\@=12

\begin{document}

\title{\Large \bf 
Unoriented Strings, Loop Equations, and $\CN=1$
  Superpotentials from Matrix Models}

\pubnum{%
RUNHETC-2002-48 \\
USC-02/08 \\
hep-th/0211291}
\date{November 2002}

\author{Sujay K. Ashok$^\flat$, Richard Corrado$^\natural$, 
Nick Halmagyi$^\natural$, Kristian D. Kennaway$^\natural$,  
and Christian R\"omelsberger$^\natural$ \\[0.4cm]
\it ${}^\flat$Department of Physics and Astronomy\\
\it Rutgers University \\
\it Piscataway, NJ 08540, USA \\[0.2cm]
\it ${}^\natural$Department of Physics and Astronomy\\
\it University of Southern California \\
\it Los Angeles, CA 90089, USA \\[0.2cm]
}

\Abstract{We apply the proposal of Dijkgraaf and Vafa to analyze
\nOne\ gauge theory 
with $\SO(N)$ and $\Sp(N)$ gauge groups with arbitrary tree-level
superpotentials using matrix model techniques. 
We derive the planar and 
leading non-planar contributions to the large $M$ $SO(M)$
and $Sp(M)$ matrix model free energy by applying the technology of
higher-genus loop equations and by straightforward 
diagrammatics. The loop equations suggest that the  $\BR\BP^2$ free
energy is
given as a derivative of the sphere contribution, a relation which we
verify diagrammatically. With a
refinement of the proposal of Dijkgraaf and Vafa for the effective
superpotential, we find agreement with field theory
expectations.}

\enlargethispage{1.5cm}

\makepapertitle

\vfill \eject 

\tableofcontents

\body


\section{Introduction}

Recently Dijkgraaf and Vafa
\cite{Dijkgraaf:2002fc,Dijkgraaf:2002vw,Dijkgraaf:2002dh} have
proposed that the exact low-energy superpotential of certain \nOne\ 
supersymmetric gauge theories is captured by the large $M$ behavior of
certain associated $M\times M$ matrix models.  This is quite
remarkable, as it 
reduces the problem of what is, in general, strongly-coupled physics
of the confining phase of pure gauge theory to the zero-dimensional
dynamics of a matrix integral.  Furthermore, the gauge theory
quantities are computed from just the planar graphs of the matrix
theory, nevertheless capturing finite $N$ results in $SU(N)$ gauge
theory.

The conjecture was initially tested for \ntwo\ $\SU(N)$ gauge theories
softly broken to \nOne\ by a tree-level superpotential for the adjoint
chiral superfields~\cite{Dijkgraaf:2002fc} and for the ${\cal N} =
1^*$ deformation of ${\cal N}=4$ $SU(N)$
SYM~\cite{Dijkgraaf:2002dh,Dorey:2002tj,Dorey:2002jc}.  The conjecture
has since been extended to a number of other
cases~\cite{Dijkgraaf:2002vw,Chekhov:2002tf,Dorey:2002pq,Ferrari:2002kq,Berenstein:2002sn,Fuji:2002wd,Gopakumar:2002wx,Demasure:2002sc,Argurio:2002xv,McGreevy:2002yg,Suzuki:2002gp,Bena:2002kw,Gorsky:2002uk,Naculich:2002hi,Tachikawa:2002wk,Dijkgraaf:2002wr,Klemm:2002pa,Dijkgraaf:2002yn,Feng:SDMM,Feng:2002yf,Argurio:2002hk,Naculich:2002hr}
and has been derived from $SU(N)$ gauge
theory~\cite{Ferrari:2002jp,Dijkgraaf:2002xd,Cachazo:2002ry}.

In this work we use matrix techniques to analyze \nOne\ gauge theory
with $\SO(N)$ and $\Sp(N)$ gauge groups.  By a careful consideration of
the planar and leading non-planar corrections to the large $M$ $SO(M)$
and $Sp(M)$ matrix
models, we derive the matrix model free energy.  We do this both
by applying the technology of higher-genus loop equations
of~\cite{Ambjorn:1993gw,Akemann:1996zr} and by straightforward
diagrammatics~(see 
{\it e.g.} \cite{Brezin:1978sv,Cicuta:1982fu}).  

The outline of the paper is as follows.  In section~\ref{sec:gauge}, we
discuss general features of the four-dimensional gauge theories with
$\CN=2$ supersymmetry softly broken to $\CN=1$.  We also suggest a
result for the superpotential of the $\CN=1^*$ theory with gauge group
$SO(2N)$ which is based on a generalization of a derivation of
Dorey~\cite{Dorey:1999sj} for $SU(N)$ gauge group.  In
section~\ref{sec:cy}, we discuss the geometric engineering of the
softly broken $\CN=2$ gauge theories by wrapping D5-branes and
O5-planes on compact cycles of generalized conifolds.  

The corresponding matrix models are introduced and solved in the large
$M$ limit in section~\ref{sec:saddle}.  As for  $SU(M)$, we find
that the loop equation for the resolvent of the matrix model describes
a Riemann surface which is identified with a factorization of the
spectral curve of the \ntwo\ gauge theory.  The large $M$ solution of
the matrix models computes the special geometry of Type IIB string
theory on the associated \CYM.

In section~\ref{sec:corrections}, we discuss the application of the
higher-genus loop equations to the computation of the $\RP^2$
contribution to the free energy.  The loop equations take the form of
integral equations which give recursion relations between the contributions to the resolvent at each
genus.  They suggest a very simple solution for the $\RP^2$
contribution in terms of the sphere contribution. In fact, the one
crosscap contribution to the resolvent $\omega_1$ satisfies
\begin{equation}
\omega_1 =\pm q\pder{\omega_0}{S_0},
\end{equation}
where $\omega_0$ is the contribution to the resolvent from the sphere.
We verify this relationship by explicitly 
enumerating several types of diagrams. We find that the contribution to
the free energy $\CF_1$ from $\BR\BP^2$ and $\CF_0$ from $S^2$ are related by
\begin{equation}
\CF_1 =\pm q \frac{\partial \CF_0}{\partial S_0},
\end{equation}
where $S_0$ is half of the 't Hooft coupling for the $SO/Sp$
component of the matrix group. We determine the proportionality
constant $q$ from the diagrammatics to be $q=\frac{g_s}{4}$.

Our results suggest a
refinement of the proposal of Dijkgraaf and Vafa for the effective
superpotential in the case of $\SO$
and $\Sp$ gauge groups. We find that
\begin{equation}
\Weff = Q_{D5}\, {\partial{\cal F}_0\over\partial S} + Q_{O5}\,\CG_0
-2\pi i \, \tau \, S,
\label{eq:DVrefine}
\end{equation}
where $Q_{D5}$ is the total charge of D5-branes, $Q_{O5}$ is the total
charge of O5-planes, $\CF_0$ is the contribution to the matrix model
free energy from diagrams with the topology of a sphere and $\CG_0$ is
proportional to $\CF_1$, the contribution to the free energy from $\BR\BP^2$
diagrams. We use~\eqref{eq:DVrefine} to obtain results
consistent with gauge theory expectations.  In particular, the matrix
model is consistent with the requirement that there is a degeneracy of
the massive vacua of the gauge theory given by $h$, the dual Coxeter
number of the gauge group.  In the case of the $\CN=1^*$ $SO(2N)$
theory, which we discuss in section~\ref{sec:onestar}, we find that
the critical value of the superpotential exactly matches the result
obtained from our gauge theory arguments in section~\ref{sec:gauge}.
We end with a discussion of our results and point out several areas
for future development. Various supporting
technical calculations are contained in appendices.

In the course of this work, two papers on matrix models with $SO/Sp$
groups have appeared. In~\cite{Fuji:2002wd}, aspects of the geometric
engineering of the gauge theories, as well the leading order in $M$
computation of the free energy of quartic orthogonal and symplectic
matrix ensembles are discussed. We use a different basis of matrices
and we account for the appearance of diagrams involving pairs of twisted
propagators that are not including in the oriented theory. More
recently, while this manuscript was in a final 
stage, \cite{Ita:Pert} appeared. These authors discuss a perturbative
derivation of the matrix model along the lines
of~\cite{Dijkgraaf:2002xd}, including a discussion of $\BR\BP^2$
corrections. Their results confirm aspects of the
refinement~\eqref{eq:DVrefine} that we found was necessary for the
computation of the gauge theory effective superpotential. 

\section{Results from \nOne\ Gauge Theories}
\label{sec:gauge}

In this section we review some results about \ntwo\ and \nOne\ 
supersymmetric gauge theories, specifically the spectral curves and
how they factorize when \ntwo\ is softly broken to \nOne.  We focus on
the case of $\SO$ and $\Sp$ gauge groups ($\SU$ was treated in
\cite{Cachazo:2001jy}, which this discussion follows).

\subsection{\ntwo\ Softly Broken to \nOne\ }

As is well known~\cite{Seiberg:1994rs,Seiberg:1994aj}\footnote{See
  \cite{Lerche:1997xu,Alvarez-Gaume:1997mv} for reviews.}, the moduli
space of \ntwo\ gauge theories is governed by a ``spectral curve'',
the periods of which give the masses of BPS objects in the theory
(W-bosons, monopoles and dyons). In
  \cite{Danielsson:1995is,Brandhuber:1995zp,Argyres:1996fw} these
  spectral curves were found for the $SO/Sp$ gauge groups. For a
  rank-$r$ gauge theory, the spectral curve is a genus $r$
  hyperelliptic curve,
\begin{equation}
y^2 = P_{2r+2}(x, \{\phi_i\}),
\label{spectral1}
\end{equation}
where $P_{2r+2}$ is a polynomial of degree $2r+2$ in the $x$ that also
depends on the moduli $\phi_i$.
The $\SO$ and $\Sp$ spectral curves can also be written as a genus
$2r-1$ curve,
\begin{equation}
y^2 = P_{2r}(x^2, \{\phi_i\}),
\label{spectral2}
\end{equation}
which is therefore symmetric under the $\Z_2$ action $x \mapsto -x$
and is a double cover of the genus $N$ curve (\ref{spectral1}) via this
map.

In \nOne\ language, the \ntwo\ vector multiplet of the Yang-Mills
theory is decomposed into an adjoint chiral superfield $\Phi$ and an
\nOne\ vector superfield $V$.  \ntwo\ \SUSY\ can be broken to
\nOne\ by an appropriate gauge-invariant superpotential term for
$\Phi$.  Because the trace of odd powers of 
matrices in the Lie algebra 
of $\SO(N)$/$\Sp(N)$ vanishes, in contrast to the $U(N)$ case
discussed in
\cite{Dijkgraaf:2002fc,Dijkgraaf:2002vw,Dijkgraaf:2002dh}, the
superpotential deformation for 
$\SO(N)$/$\Sp(N)$ only includes polynomial terms of even
degree:
\begin{equation}
\Wtree(\Phi) = \sum_{k=1}^{n+1} \frac{g_k}{2k} \hbox{Tr}(\Phi^{2k}).
\label{wtree}
\end{equation}
A superpotential $\Wtree$ of order $2n+2$ breaks the gauge symmetry down
to a direct product of $n+1$ subgroups, {\it e.g.}:
\begin{equation}
\SO(N) \rightarrow \SO(N_0) \times U(N_1) \times \ldots \times U(N_n),
\end{equation} 
where $N=N_0+2N_1+\cdots+2N_n$.

The $U(1)$ factors in this theory decouple in the IR. In the
supersymmetric vacua of the \nOne\ theory $r-n$ mutually local
monopoles simultaneously become massless and condense, leading to
confinement of the gauge theory \cite{Cachazo:2001jy}. The condition
that $r-n$ mutually local monopoles become massless leads to a
``factorization locus'' in the moduli space of the spectral curve, where
$r-n$ of the (non-intersecting) cycles of the spectral curve are
simultaneously pinching off to zero-volume.   

Imposing this condition is therefore equivalent to the
factorization~\cite{Cachazo:2002pr} 
\begin{equation}
y^2 = \prod_{i=1}^{r-n} (x^2 - p_i^2)^2 \prod_{j=1}^{2n}(x^2 - q_i^2),
\end{equation}
where $p_i \neq p_j, q_i \neq q_j$ for $i \neq j$.  On this locus we
then obtain the ``reduced spectral curve''
\begin{equation}
y^2 = \prod_{j=1}^{2n}(x^2 - q_i^2),
\label{n1curve}
\end{equation}
which is a genus $2n-1$ curve.  This curve parameterizes the \ntwo\ 
vacua that are not lifted by the deformation to \nOne\ (\ref{wtree}).
Notice that the factorized curves now have a similar form for $\SO$ and
$\Sp$, and the curve is still invariant under $x \mapsto -x$ (this
implies that the branch points come in pairs: $(-q_i, q_i)$).  This
reduced spectral curve will be derived from string theory in the
following section by taking an orientifold action on the configuration of
D-branes on a
generalized conifold that engineers this \nOne\ gauge theory.

The low-energy effective superpotential of these gauge theory can be
obtained from the reduced spectral curve as discussed
by~\cite{Cachazo:2001jy,Cachazo:2002pr}. It will take the form
\begin{equation}
\Weff = \sum_i \left( \hat{N_i} \Pi_i - 2 \pi i \tau_i S_i \right),
\label{weff}
\end{equation}
where $2 \pi i S_i$ are the periods of the meromorphic 1-form $y\,dx$
around the A-cycles of the spectral curve, $\Pi_i$ the
corresponding periods 
around the B-cycles, and $\hat{N_i}$ is  
\begin{equation}
\hat{N_i} = \left\{\begin{array}{cc}N_i & \SU(N_i), \\
\frac{N_i}{2} - 1 & \SO(N_i), \\
N_i + 1 & \Sp(N_i).
\end{array}\right.
\end{equation}
By contrast,
we will find that the shift $N_i \mapsto \hat{N_i}$ emerges in the
matrix model by considering the first subleading corrections to the
large $M$ expansion, coming from Feynman diagrams of topology $\RP^2$.

Recently the gaugino effective superpotential has been perturbatively
derived from $SU(N)$ gauge
theory~\cite{Ferrari:2002jp,Dijkgraaf:2002xd,Cachazo:2002ry}.  It is
interesting to note that similar arguments\footnote{We thank Jaume
  Gomis and Jongwon Park for discussions on this issue. The same
observation about crosscap contributions was made in~
\cite{Ita:Pert} and more recently for boundaries
in~\cite{Bena:2002ua}.} to those  
of~\cite{Dijkgraaf:2002xd} can be used to argue that only diagrams
with at most one boundary (if quark flavors are present) or crosscap
will contribute to the gauge theory
superpotential~\cite{Dijkgraaf:2002dh,Argurio:2002xv,Bena:2002kw}.

\subsection{The $\CN=1^*$ Theories}

The $\CN=1^*$ theories arise as deformations of  $\CN=4$ Yang-Mills
theory by mass terms for the three adjoint $\CN=1$ chiral fields. The
total superpotential is
\begin{equation}
W= \tr\left( \Phi_1[\Phi_2,\Phi_3] + \sum_{i=1}^3 m_i\, \Phi_i^2 \right)
\end{equation}
and the F-flatness conditions can be written as
\begin{equation}
[\Phi_i,\Phi_j] \propto i\epsilon_{ijk}  \Phi_k.
\end{equation}
Supersymmetric vacua are then obtained by embedding $SU(2)$
representations into the gauge group $G$. In particular, for
$G=SU(N)$, the embeddings are classified by the divisors $d$ of $N$, leading to  $\sum_{d|N} d$ massive
vacua~\cite{Donagi:1996cf}. 

Dorey~\cite{Dorey:1999sj} (see also~\cite{Dorey:2000fc}), following
the approach of~\cite{Seiberg:1996nz}, 
compactified the theory with $SU(N)$ gauge group on a circle of
radius~$R$. The degrees of freedom of the effective $2+1$--dimensional
theory are $r={\rm rank}(G)$ complex Abelian scalar fields $X_a$ which
are composed of the Wilson lines and the scalars dual to the massless
photons of the theory. The moduli space of the theory is 
\begin{equation}
\CM = E_\tau^r/\CW_G,
\end{equation}
where $E_\tau$ is the elliptic curve parameterized by each $X_a$ and
$\CW_G$ is the Weyl group of $G$.  

By several arguments, including the relationship
between the elliptic 
Calogero-Moser systems and the $\CN=2$ theories which can be softly
broken to the $\CN=1^*$ theory, Dorey found that the superpotential of
the $2+1$--dimensional theory took the form
\begin{equation}
W = c \sum_{a>b}  \wp(X_a-X_b), \label{eq:sunweier}
\end{equation}
where $\wp(z)$ is the Weierstrass function. Dorey argued that the
coefficient $c$ was independent of the radius $R$, so
that critical values of~\eqref{eq:sunweier} (which depend on the
modular parameter $\tau$ and not the $X_a$) could be evaluated in the
vacua of the theory and extrapolated directly to the
$R\rightarrow\infty$ limit.

It is interesting to ask what the generalization\footnote{After an
initial version of this paper appeared, we became aware of earlier
work by Kumar and Troost~\cite{Kumar:2001iu}, where they also suggest the
formula~\eqref{eq:gweier} and give many more arguments for its
validity. We thank P.~Kumar for making us aware of this.}
of~\eqref{eq:sunweier} is to arbitrary gauge groups~$G$. The modular
properties of the superpotential that were crucial to Dorey's argument
must still be preserved, so the superpotential should remain a sum of
Weierstrass functions. An obvious guess for the argument of these
functions is to replace $X_a-X_b$ by the sum over the positive roots
$\sum_{\alpha>0}\alpha\cdot X $. The integrable systems
approach~\cite{Gorsky:1995zq,Martinec:1996by,Donagi:1996cf,Martinec:1996qn} to 
$\CN=2$ theories is a promising route to this result. In fact, D'Hoker
and Phong~\cite{DHoker:1998ha,DHoker:1998yg,DHoker:1998yh} have 
determined the integrable systems that govern a large class of
$\CN=2$ theories
with gauge group $G$. An application of the techniques
of~\cite{Donagi:1996cf,Dorey:1999sj} to the soft breaking of these
theories to  $\CN=1^*$ suggests that the correct superpotential for
gauge group $G$ is
\begin{equation}
W = c \left(
\sum_{\{\alpha_L>0\}}\wp(\alpha_L\cdot X )
+\sum_{\{\alpha_S>0\}}
\wp_{\nu(\alpha_S)}(\alpha_S\cdot X)
 \right).
\label{eq:gweier}
\end{equation}
where $\alpha_{L,S}$ are the long and short positive roots of the Lie
algebra of $G$,
respectively, and $\wp_\nu(z)$ are the {\it twisted} Weierstrass
functions 
\begin{equation}
\wp_\nu(z) = \sum_{\sigma=0}^{\nu-1}
\wp\left(z+2\omega_a \frac{\sigma}{\nu}\right) .
\end{equation}
defined in~\cite{DHoker:1998ha}. For
non-simply laced groups, roots of only two different lengths appear:
$\nu(\alpha)=1$ for all long roots, $\nu(\alpha)=2$ for all short
roots of $\frak{b}_n$, $\frak{c}_n$ $\frak{f}_4$, while
$\nu(\alpha)=3$ for the short roots of $\frak{g}_2$.

For $SO(2N)$, since it is simply-laced, twisted Weierstrass
functions do not appear, and we obtain 
\begin{equation}
W = c \sum_{a>b} \left[ \wp(X_a-X_b)+\wp(X_a+X_b)\right]. 
\label{eq:soevenweier}
\end{equation}
Following~\cite{Dorey:1999sj}, we can evaluate this in the $k^{\rm th}$ 
confining vacuum to find (up to an additive constant)
\begin{equation}
W \sim E_2\bigl((\tau+k)/(2N-2)\bigr),
\label{eq:soevenWcrit}  
\end{equation}
where $E_2(\tau)$ is the second regularized Eisenstein series. It is
tempting to conjecture that the result for arbitrary $G$ will take
this form with the obvious substitution of $h$, the dual Coxeter number
of $G$ for $2N-2$, but this remains to be verified. 

\section{Calabi-Yau Geometry}
\label{sec:cy}

We now review the string theoretic engineering of a softly broken
$\CN=2$ gauge theory with $SO/Sp$ gauge group
\cite{Cachazo:2002pr,Cachazo:2001jy,Edelstein:2001mw,Dijkgraaf:2002fc}.
We consider type IIB string theory compactified on the non-compact
$A_1$ fibration
\begin{equation}
u^2 + v^2 + w^2 + W'(x)^2 = 0,
\end{equation}
where $W(x)$ is a degree $n+1$ polynomial, which will later be
related to the tree level superpotential.  This fibration has
singularities at the critical points of $W(x)$.  In the neighborhood
of those singularities, we can introduce the coordinate $x'=W'(x)$.
Then it is easy to see that the singularities are all conifold
singularities.

This generalized conifold can be de-singularized in two ways: it can be
resolved or it can be deformed.  The resolution is given by the
surface
\begin{equation}\label{resgenconi}
\left(\begin{array}{cc} u+iv & w+iW'(x) \\ -w+iW'(x) & u-iv \end{array}\right)
\left(\begin{array}{c} \lambda_1 \\ \lambda_2 \end{array}\right)=0
\end{equation}
in $\BC^4\times\BP^1$.  In this geometry each singular point is
replaced by a $\BP^1$.  These $\BP^1$'s are disjoint, holomorphic,
have the same volume and are homologically equivalent.  The latter
property can be seen by making use of the fibration structure away
from $W'(x)=0$.  This $A_1$ fibration over the $x$ plane induces a
fibration of non-holomorphic $S^2$'s over the $x$ plane.  This $S^2$
cannot shrink to zero size as one approaches a critical point of $W$
in the $x$ plane, but it becomes the holomorphic $\BP^1$ of the
resolution.

We can now construct a softly broken $\CN=2$ $U(N)$ gauge theory with
tree level superpotential $W(x)$ by wrapping $N$ D5-branes around the
$S^2$.  This is an UV definition of the theory.  A classical
supersymmetric vacuum is obtained by minimizing the volume of the
D5-branes. This amounts to distributing a collection of $N_i$
D5-branes over the $n$ minimal-volume holomorphic $\P^1$'s at the
critical points of $W$.  The $U(N)$ gauge symmetry is then
spontaneously broken to $U(N_1)\times\cdots\times U(N_{n-1})$.

We now want to consider an orientifold of this
theory\footnote{Orientifolds were discussed in the A-model  
in~\cite{Sinha:2000ap,Edelstein:2001mw,Acharya:2002ag,Fuji:2002vv},
while the discussion of~\cite{Gomis:2001xw} is more closely related to
the B-model which is our interest here.}.  Since we 
started with a type IIB theory on a Calabi Yau, we have to combine the
worldsheet orientation reversal with a holomorphic involution of the
\CY\ (an anti-holomorphic involution would be appropriate for the IIA
theory).  Furthermore we want to fix
one of the $\BP^1$'s and act freely on the rest of the Calabi Yau
geometry.  This can be done if $W(x)$ is an even polynomial of order
$2n$.  In terms of the fibration structure of the \CY\, this means
that the critical points of $W'(x)$ come in pairs $(-x_i, x_i)$ and \
one critical point is fixed at $x_0=0$. Then
\begin{equation}\label{holoinv}
(u,v,w,x,\lambda_1,\lambda_2)\mapsto(-u,-v,-w,-x,\lambda_1,\lambda_2)
\end{equation}
is a holomorphic involution of the geometry (\ref{resgenconi}), which
leaves only the $\BP^1$ at $u=v=w=x=0$ fixed.  In the string
theory this means that there is an O5-plane wrapping this $\P^1$ in
the \CY\ geometry.  

There are essentially two choices of O5-plane with which we can wrap
the fixed $\P^1$.  They are distinguished by a different choice of
worldsheet action and carry RR 5-form charge of $\pm 1$ (the RR charge
of an O$p^\pm$-plane is $\pm 2^{p-5}$ in conventions where we count the
charge of $N/2$ D-branes but not their $N/2$
images). 
The orientifold contribution to 
the RR charge of objects wrapping the $\P^1$ will cause a shift in the
coefficient $N_0$ in the flux-generated superpotential on the deformed
\CY\ geometry, as explained below.

Now we can construct a softly broken $\CN=2$ $SO(N)/Sp(N/2)$ gauge
theory with tree level superpotential $W(x)$ by wrapping $N$ D5-branes
around the $S^2$ and then performing the orientifold.  The gauge symmetry is again broken $SO(N) \mapsto
SO(N_0) \times U(N_1) \times \cdots \times U(N_{n-1})$ or $Sp(N/2)
\mapsto Sp(N_0/2) \times U(N_1) \times \cdots \times U(N_{n-1})$
respectively with $N=N_0+2N_1+\cdots+2N_{n-1}$.

If we flow this ultraviolet theory to the infrared, there will be a
confinement transition.  In string theory this is described by a
geometric transition in which the resolved conifold geometry with
wrapped D5-branes and O5-planes is replaced by a deformed conifold
geometry~\cite{Vafa:2000wi}
\begin{equation}
u^2 + v^2 + w^2 + W'(x)^2 - f(x) = 0,
\end{equation}
where $f(x)$ is an even polynomial of degree $2n-2$.  Such a
polynomial represents the most general normalizable deformation of the
singular conifold that still respects the holomorphic involution
(\ref{holoinv}).  For a reasonably small $f(x)$, each critical point
of $W'(x)$ is replaced by two simple zeros of $W'(x)^2 - f(x)$.  This
means that each $\BP^1_i$ is replaced by a 3-sphere $A_i$ with 3-form
RR-flux $H$ through it, equal to the amount of D5-brane and O5-plane
charge on the $\BP^1_i$.  The orientifold acts on one 3-sphere $A_0$
as the antipodal map, while the other 3-spheres are mapped to each
other in pairs $A_i$ and $A_{-i}$. Note that 
there is no orientifold fixed plane anymore.

The coefficients in $f(x)$ are normalizable modes that are localized
close to the tip of the conifold.  The coefficients in $f(x)$ are
determined by the periods
\begin{equation}
S_i=\frac{1}{2\pi i}\int_{A_i}\Omega.
\end{equation}
These periods $S_i$ can be interpreted as the gaugino condensates of
the gauge theory.  There are non-compact 3-cycles $B_i$ that are dual
to the $A_i$.  The periods of the B-cycles are
\begin{equation}
\pder{\CF_0}{S_i}=\int_{B_i}\Omega,
\end{equation}
where $\CF_0$ is the prepotential.  One needs to introduce a cutoff in
order to make these periods finite.

The flux through the cycles $A_i$ is determined in terms of the
RR-charges of the D-brane and O-plane configuration
\begin{equation}
\begin{split}
N_0\pm 2&=\int_{A_0}H,\\
N_i&=\int_{A_i}H, ~~~i\ne 0,
\end{split}
\end{equation}
and the flux through the cycles $B_i$ is given in terms of the
coupling constants
\begin{equation}
\tau_i=\int_{B_i}H.
\end{equation}
Since there is no orientifold fixed plane, there are no contributions
 to the effective superpotential for the
gaugino condensate from unoriented closed
 strings~\cite{Acharya:2002ag}.  It is then given by the 
flux superpotential
\cite{Taylor:1999ii,Becker:1996gj,Gukov:1999ya,Polchinski:1996sm}
\begin{equation}
W_{eff}(S_i)=\int H\wedge\Omega,
\end{equation}
where the integral is taken only over half of the covering space of
the orientifold. 
Using the expressions for the periods and the fluxes and taking into
account the orientifold projection, we get
\begin{equation}
\Weff(S_i)=
\left(\frac{N_0}{2}\pm 1\right)\pder{\CF_0}{S_0}
+\sum_{i>0}N_i\pder{\CF_0}{S_i}-
\half\tau_0S_0-\sum_{i>0}\tau_iS_i. \label{eq:weffclosed}
\end{equation}

This result could also have been computed on the open string side
before the transition.  On the open string side there is no flux
through any 3-cycles, so there is no contribution to the
superpotential due to closed oriented strings.  But there are two
kinds of other contributions to the effective superpotential: the open
string contributions (disk diagrams) and the contributions due to
closed unoriented strings at the orientifold fixed plane ($\RP^2$
diagrams). The contribution due to the open strings is the equal to
one half that of the theory without the orientifold, {\it i.e.}, it is
\begin{equation}
\Weff^{o}(S_i)=\frac{N_0}{2}\pder{\CF_0}{S_0}
+\sum_{i>0}N_i\pder{\CF_0}{S_i}-
\half\tau_0S_0-\sum_{i>0}\tau_iS_i.
\end{equation}
The contribution due to the unoriented closed strings then must  be
\begin{equation}
\Weff^{u}(S_i)=W_{eff}(S_i)-W^{o}_{eff}(S_i)=\pm\pder{\CF_0}{S_0}.
\end{equation}
We will confirm this result in our matrix model computation.

\section{The Classical Loop Equation}
\label{sec:saddle}

We first consider the saddle point evaluation of the one matrix
integral for $SO(M)$ or $Sp(M)$ matrices. Our discussion is analogous to
that of \cite{DiFrancesco:1995nw,Dijkgraaf:2002fc} and consists of
obtaining a loop equation for the resolvent. In the next section, we
will formulate a systematic method to obtain the $g_s$ corrections to the
classical solution.

The partition function for the model with one matrix $\Phi$ in the
adjoint representation of the Lie algebra of $G=SO(M)$ or $Sp(M)$ is
\begin{equation}
Z = {1\over {\rm Vol}\, G} \int d\Phi \, 
\exp \left( - {1\over g_s} {\rm Tr}\, W(\Phi)\right).
\label{eq:onematrix}
\end{equation}
In Appendix~\ref{appvand}, we collect results that are useful for
$SO/Sp$ groups, but here we shall discuss only the $SO(2M)$  group
in detail.  

In the eigenvalue basis, the integral over an $SO(2M)$ matrix is given
by  
\begin{equation}
Z\sim\int\prod_{i=1}^M \, d\lambda_{i}\,
\prod_{i<j} (\lambda_i^2-\lambda_j^2)^2 \,
e^{-\frac{2}{g_s}\sum_i W(\lambda_i)}.
\end{equation}
The effective action for the gas of eigenvalues is given by
\begin{equation}
S(\lambda)=-\sum_{i<j}\ln(\lambda_i^2-\lambda_j^2)^2+{2\over g_s}
\sum_i W(\lambda_i).
\end{equation}
Note that $W$ is now a polynomial of order $2n$ with only even powers. This 
is because the trace of an antisymmetric matrix vanishes. 
In principle $W(\Phi)$ could also contain the Pfaffian, 
but we will omit this case.

This action gives rise to the classical equations of motion
\begin{equation}
\sum_{j\ne i}{2\lambda_i\over\lambda_i^2-\lambda_j^2}
-{1\over g_s}W'(\lambda_i) =0.
\end{equation}
It is useful to define the resolvent
\begin{equation}
\omega_{0}(x)=g_s\Tr{1\over{x-\Phi}}=
g_s\sum_i{2x\over{x^2-\lambda_i^2}},
\end{equation}
then, by multiplying the equations of motion by 
$\frac{2\lambda_i}{x^2-\lambda_i^2}$ and summing over $i$, we obtain
an equation for $\omega_{0}(x)$ exactly as for $SU(M)$: 
\begin{equation}
\omega_0(x)^2-g_s\left(\frac{\omega_0(x)}{x}-\omega_0'(x)\right)
+f(x)-2\omega_0(x)W'(x) =0, \label{eq:eomresolvent}
\end{equation}
where
\begin{equation}\label{fofx}
f(x)=g_s\sum_i\frac{2\lambda_iW'(\lambda_i)-2xW'(x)}{\lambda_i^2-x^2}
\end{equation}
is a polynomial of order $2n-2$ with only even powers, {\it i.e.}, it
has $n$ coefficients.

In the small $g_s$ limit, \eqref{eq:eomresolvent} reduces to
\begin{equation}
\omega_0(x)^2+f(x)-2\omega_0(x)W'(x)=0,
\end{equation}
or
\begin{equation}\label{spectral}
y^2-W'(x)^2+f(x)=0,
\end{equation}
where
\begin{equation}
y(x)=\omega_0(x)-W'(x).
\end{equation}
The force equation is then
\begin{equation}\label{forceeqn}
2y(\lambda)=-g_s{\partial S\over\partial\lambda},
\end{equation}
where the factor of 2 comes from the fact that the force is acting on
an eigenvalue and its image. This is the same equation as for $SU(2M)$,
with the only  
difference being that the polynomials $W$ and $f$ have only even powers.
This matches the expected result from the orientifold procedure in string
theory. 

The equation for the resolvent can be solved using (\ref{spectral}), 
yielding a formal solution~\cite{DiFrancesco:1995nw} 
\begin{equation}
\omega_{0}(x)=W'(x)-\sqrt{W'(x)^2-f(x)}.
\end{equation}
The resolvent is thus expressed in terms of the $n$ unknown coefficients
that appear in the polynomial $f(x)$ defined in
(\ref{fofx}). From the 
form of the solution, it is clear that the resolvent has branch cuts
among 
which the eigenvalues of the matrix are distributed. In the large $M$
limit, we thus get a distribution of eigenvalues, with the eigenvalue
density given by
$\rho(\lambda)$  
\begin{equation}
\omega_0(x)=2\int_0^\infty{x\rho(\lambda)d\lambda\over{x^2-\lambda^2}}=
\int_0^\infty\rho(\lambda)d\lambda\left({1\over{x-\lambda}}+
{1\over{x+\lambda}}\right)=
\int_{-\infty}^\infty{\rho(\lambda)d\lambda\over{x-\lambda}},
\end{equation}
which implies that
\begin{equation}
\rho(\lambda)={1\over 2\pi i}(\omega_0(\lambda+i0)-\omega_0(\lambda-i0))=
{1\over 2\pi i}(y(\lambda+i0)-y(\lambda-i0)).
\end{equation}

The filling fractions are then given by
\begin{equation}
\begin{split}
S_0&={1\over 4\pi i}\int_{A_0}y(x)dx,\\
S_i&={1\over 2\pi i}\int_{A_i}y(x)dx~~ ,i>0
\end{split}
\end{equation}
Note that we only integrate around half of the cycle $A_0$ because of
the orientifold projection.
At the classical level, one can see from (\ref{forceeqn}) that $y(x)$ is
the force acting on an eigenvalue. Now, the variation of the free energy
$\CF_0$ of the matrix model caused by a changing the number of
eigenvalues on the $i^\text{th}$ cut is then the line integral of the
force over 
the non compact $B_i$ cycle of the Riemann surface (\ref{spectral})
 \begin{equation}
{\partial{\cal F}_0\over\partial S_i}=\int_{B_i}y(x)dx.
\end{equation}
This is the differential equation that determines ${\cal F}_0$, 
{\it i.e.}, the 
leading contribution to the free energy.  

For $SO(2M+1)$ and $Sp(M)$, one can easily see that $\CF_0$ and the
Riemann surface are the same as in the case of $SO(2M)$.  
In the next section we will determine the leading contribution
from unoriented diagrams to the free energy, which is a subleading term in
the $g_s$ expansion of the free energy.

\section{$g_s$ Corrections and  Loop Equations}
\label{sec:corrections}

The partition function of the $SO/Sp$ matrix model is
\begin{equation}\label{matrint}
Z = e^{\frac{1}{g_s^2}\CF} = \int d\Phi e^{-\frac{1}{g_s}\Tr W(\Phi)},
\end{equation}
where the overall coupling constant $g_s$ can be thought of as the
string coupling and the action is 
\begin{equation}
W(\Phi)=\sum_{j=1}^{\infty}\frac{g_{j}}{2j}\Phi^{2j}.
\end{equation}
Dijkgraaf and Vafa
\cite{Dijkgraaf:2002fc,Dijkgraaf:2002vw,Dijkgraaf:2002dh} conjectured
that the exact superpotential of the gauge theory with the tree level
superpotential $W(\Phi)$ is given by the perturbative expansion of the
matrix integral (\ref{matrint}) around one classical vacuum (saddle
point). Such a classical vacuum is given by a distribution of the
eigenvalues of $\Phi$ over the critical points $\{x_i\}$ of the
superpotential $W(x)$. We denote the number of eigenvalues at the critical
point $x_i$ by $M_i$ and define the parameters
\begin{equation}
S_0 = g_s \frac{M_0}{2},~~S_i=g_s M_i.
\end{equation}
The perturbative expansion around such a classical vacuum can be
visualized in terms of fat graphs, where edges of a ribbon correspond to
Chan-Paton factors. For each Chan-Paton factor we have to choose a
critical point $x_i$, on which it sits and a loop of such a Chan-Paton
factor gives a contribution of $M_i=S_i/g_s$. From the overall
normalization of the action, it is clear that each vertex of the diagram
contributes a factor of $1/g_s$ and each propagator contributes $g_s$.
Thus the overall power of $g_s$ counts the Euler character of the fat
graph. The superpotential is then given only by the contributions from
planar and $\RP^2$ diagrams. These diagrams have a very simple $g_s$
dependence, but the $S_i$ dependence can actually be quite complicated, as
we will see.

If one did the full matrix integral, there would be a sum over all saddle
points and the $S_i$ dependence would be lost. However, since we are
interested in only in a perturbative expansion around a classical vacuum,
the $S_i$ dependence is nontrivial and will describe how the effective
superpotential depends on the gaugino condensates.

In a recent paper \cite{Dijkgraaf:2002pp}, it has been shown that the in a
vacuum where the gauge symmetry is broken to a subgroup (say, a product of
$U(N_i)$ factors), the off diagonal components of the matrix $\Phi$ do not
correspond to propagating degrees of freedom, and that these should be
properly interpreted as the Faddeev-Popov ghosts that are necessarily
included because of the gauge fixing involved in doing the matrix model.
Thus, for computing Feynman diagrams in the matrix model, we have to
include terms in the Lagrangian that belong to the ghost sector as well.
But the loop equations, as we shall see, correspond to Ward identities in
the matrix model. They arise because of the invariance of the matrix
integral under an arbitrary reparametrization of $\Phi$ that respects the
$SO/Sp$ symmetry of the Lagrangian. If we take into account the variation
of the measure as well, then this symmetry leads to the loop equations.
Thus, we do not expect the ghosts to be relevant for the discussion in
this section.

In the $SO/Sp$ case we expand the matrix model partition function in a
systematic expansion in $g_s$. The coefficients of the terms in the
expansion are the contributions coming from the Feynman graphs that can be
drawn on a surface of Euler character $\chi=2-2g-c$ where $g$ denotes the
genus, and $c$ denotes the number of cross-caps. We mentioned earlier
that each loop in a Feynman diagram contributes a factor $M$. In order to
see this, consider the propagator for the $SO(M)$ matrix model. It has a
group theoretic factor
\begin{equation}
\langle 
\Phi_{ij}\Phi_{kl}\rangle\sim\half(\delta_{ik}\delta_{jl}-\delta_{il}\delta_{jk}).
\end{equation}
Thus, each loop in a Feynman diagram contributes a factor of $M_i$ (the
number of Chan-Paton factors on the $i$th critical point). 

\subsection{The Resolvent}

We shall now introduce the general technique of loop equations, which is
an iterative procedure to calculate the higher order (in $g_s$)
corrections to the partition function. Central to this procedure is the
loop operator defined as
\begin{equation}
\frac{d}{dV}(x) = -\sum_{j=1}^{\infty}\frac{2j}{x^{2j+1}}\pder{}{g_{j}}.
\end{equation}
The resolvent, which is the generating functional for the single trace
correlation functions of the matrix model is defined as
\begin{equation}
\omega(x)=g_s\left\langle\Tr\frac{1}{x-\Phi}\right\rangle = g_s\sum_{k=0}^{\infty}\frac{\langle\Tr\Phi^{2k}\rangle}{x^{2k+1}}
\end{equation} 
Using the identity 
\begin{equation}
-(2k)\frac{d}{dg_{k}}\CF=g_s\langle\Tr\Phi^{2k}\rangle,
\end{equation}
we can express the resolvent as
\begin{equation}\label{dFbydV}
\omega(x) = \frac{d}{dV}(x)\CF + \frac{S}{x},
\end{equation}
where we used $S=\sum S_i=g_s M$. We are using the variables $g_s$ and
$S$, since we are working in the small $g_s$ limit with $S$ fixed. As
mentioned before, the free energy has an expansion in $g_s$ of the form
\begin{equation}
\CF = \sum_{g,c}g_s^{2g+c}\CF_{g,c}
\end{equation}
We will be interested in calculating the first two terms in this
expansion, which are the contributions from diagrams with the topology of
$S^2$ and $\RP^2$. The resolvent has a similar expansion 
\begin{equation}
\omega(x) = \sum_{g,c}g_s^{2g+c}\omega_{g,c}(x).
\end{equation}

The asymptotic behavior at infinity of the $\omega_{g,c}$ is clear from
the definition of $\omega(x)$
\begin{equation}\label{Wasympt}
\begin{split}
\omega_{0,0}(x)=&\frac{S}{x}+\CO(x^{-2}),\\
\omega_{g,c}(x)=& \CO(x^{-2}),\;\;\;\;2g+c>0.
\end{split}
\end{equation}
Using this fact and the existence of the genus expansion, we can write 
\begin{equation}\label{wfromf}
\begin{split}
\omega_{0,0}(x) =& \frac{d}{dV}(x)\CF_{0,0}+ \frac{S}{x},\\
\omega_{g,c}(x) =& \frac{d}{dV}(x)\CF_{g,c},\;\;\;\;2g+c>0.
\end{split}
\end{equation}

These equations determine the dependence of $\CF_{g,c}$ on the coupling
constants. There is still an additive constant which is undetermined,
but this is unphysical. In the next section we will derive the loop
equation, which will provide us with recursion relations to calculate
$\omega_{g,c}$ as functions of the coupling constants $g_j$. (For the rest
of the discussion, we denote $\omega_{0,0}$ by $\omega_0$ and
$\omega_{0,1}$ by $\omega_1$.)

\subsection{The Loop Equation}

In this section we will derive an important recursion relation between
the different perturbative contributions $\omega_{g,c}$ to the
resolvent\footnote{In an earlier version of this paper, the last term
in~\eqref{eq:phijacobian} was missed. This was corrected
in~\cite{Janik:2002nz}. While our result for the $\RP^2$
contribution~\eqref{foneansatz} is unchanged, we have corrected our
derivation here. The same result was obtained via different methods
in~\cite{Janik:2002nz}.}. 
The loop equation can be derived by doing a reparametrization of the
matrices $\Phi$ in the matrix integral and observing that the integral
is invariant under this reparametrization. Let us reparametrize $\Phi$
by 
\begin{eqnarray}\label{reparam}
\Phi&=&\Phi'-\left(\frac{\epsilon}{x-\Phi'}\right)_\odd=
\Phi'-\epsilon\sum_{k=0}^\infty\frac{\Phi'^{2k+1}}{x^{2k+2}}\\
d\Phi&=&d\Phi'-\epsilon\sum_{k=0}^\infty
\sum_{l=0}^{2k}\frac{\Phi'^ld\Phi'\Phi'^{2k-l}}{x^{2k+2}}
\end{eqnarray}
where we only take the odd/even powers of $\Phi'$ in order to preserve
the $SO/Sp$ Lie algebra. The Jacobian for this reparametrization is
then 
\begin{equation}
J(\Phi')=1-\frac{\epsilon}{2}\left(\Tr\frac{1}{x-\Phi'}\right)^2
+\frac{\epsilon}{2x} \Tr\frac{1}{x-\Phi'}.  
\label{eq:phijacobian}
\end{equation}
The action transforms as
\begin{equation}
\Tr W(\Phi)=
\Tr W\left(\Phi'-\left(\frac{\epsilon}{x-\Phi'}\right)_\odd\right)
= \Tr W(\Phi')-\epsilon\, \Tr\frac{W'(\Phi')}{x-\Phi'}.
\end{equation}
Inserting this into the matrix integral, we get
\begin{equation}
\begin{split}
& \half\int
d\Phi'\left[\left(\Tr\frac{1}{x-\Phi'}\right)^2
-\frac{1}{x}\Tr\frac{1}{x-\Phi'}\right]e^{-\frac{1}{g_s}\Tr
W(\Phi')} \\
& \hskip2cm =
\frac{1}{g_s}\int d\Phi'\Tr\frac{W'(\Phi')}{x-\Phi'}
e^{-\frac{1}{g_s}\Tr W(\Phi')}.
\end{split}
\end{equation}
We can now make use of the identity
\begin{equation}
\frac{d}{dV}(x)\omega(x)=
\left\langle\left(\Tr\frac{1}{x-\Phi}\right)^2\right\rangle-
\left\langle\Tr\frac{1}{x-\Phi}\right\rangle^2
\end{equation}
to get the loop equation
\begin{equation}
g_s\left\langle\Tr\frac{W'(\Phi)}{x-\Phi}\right\rangle=
\half\omega(x)^2-\frac{g_s}{2x}\omega(x)
+\frac{g_s^2}{2}\frac{d}{dV}(x)\omega(x).
\end{equation}

We can rewrite the loop equation using
\begin{equation}
g_s\left\langle\Tr\frac{W'(\Phi)}{x-\Phi}\right\rangle=
g_s\left\langle\sum_{i}\frac{W'(\lambda_{i})}{x-\lambda_{i}}\right\rangle
=\oint_{C}\frac{dx'}{2\pi i}\frac{W'(x')}{x-x'}\omega(x'),
\end{equation}
where $C$ is a contour that encloses all the eigenvalues of $\Phi$ but not $x$.
In the small $g_s$ (large $M$) limit of the matrix model, we get a
continuous eigenvalue distribution for $\Phi$ and all the eigenvalues
are distributed over cuts on the real axis of the $x$-plane. The loop
equation now reads 
\begin{equation}\label{loopeqn}
\oint_{C}\frac{dx'}{2\pi i}\frac{W'(x')}{x-x'}\omega(x') = 
\half\omega(x)^2-\frac{g_s}{2x}\omega(x)
+\frac{g_s^2}{2}\frac{d}{dV}(x)\omega(x).
\end{equation}
We can now insert the $g_s$ expansions for the resolvent and
iteratively solve for the $\omega_{g,c}$. The zeroth and first order
equations are 
\begin{eqnarray}\label{loworderloop}
\oint_{C}\frac{dx'}{2\pi i}\frac{W'(x')}{x-x'}\omega_{0}(x') &=& 
\half\omega_{0}(x)^2, \\\label{loworderloop2}
\oint_{C}\frac{dx'}{2\pi i}\frac{W'(x')}{x-x'}\omega_{1}(x') &=& 
\omega_{0}(x)\omega_{1}(x)-\frac{1}{2x}\omega_0(x).
\end{eqnarray}

The resolvent that solves the loop equations has to satisfy
(\ref{Wasympt}) which imposes constraints on the end points of the
cuts in the $x$-plane. Note that this derivation of the loop equation
is valid in the saddle point approximation that we are using. 

Equation \eqref{loworderloop2} is a linear inhomogenous integral
equation for $\omega_1$. The homogeneous equation is solved by a
derivative of $\omega_0$ with respect to any parameter which
specifies the vacuum, {\it i.e.}, is independent of the coupling
constants $g_j$. In our case there are only the parameters $S_i$,
which specify the classical vacuum around which the matrix integral is
expanded. We will elaborate on this observation in section
\ref{sec:rptwocontr} for the case of a softly broken \ntwo\ theory and
we will use this result for the $\CN=1^*$ theory. 

\subsection{Solution to the Loop Equations}

Let us now solve the the loop equations (\ref{loworderloop}) first for $\omega_0$ and then for $\omega_1$ in the case of a polynomial potential
\begin{equation}
W(\Phi)=\sum_{j=1}^{n}\frac{g_j}{2j}\Phi^{2j}.
\end{equation}
In this section, we closely follow the discussion in  
\cite{Ambjorn:1993gw,Akemann:1996zr}.

\subsubsection{Planar Contributions}

In equation (\ref{loworderloop}), we deform the integration contour
$C$ to encircle infinity, and rewrite it as 
\begin{equation}
\half\omega_0(x)^2 
=W'(x)\omega_0(x)
+\cint{C_{\infty}}{x'}\frac{W'(x')\omega_{0}(x')}{x-x'}.
\end{equation}
Assuming that $\omega_0(x)$ has $k$ cuts in the complex $x$-plane, we
make the ansatz  
\begin{equation}
\omega_0(x)=W'(x)-M(x)\sqrt{\prod_{i=1}^{2k}(x-x_i)},
\end{equation}
where $M(x)$ is an undetermined analytic function at the moment. Here the
end points of the cuts, denoted by the $x_i$, are unknown and have to be
determined. From the discussion in \cite{Akemann:1996zr}, it is clear that
if we have the maximum number $k=2n-1$ of cuts  allowed, the function
$M(x)$ is a constant. The loop equation determines $M$ in this case to be
the coupling constant $g_n$. Also, in the $SO/Sp$ case the eigenvalues
come in pairs and the total number of ``independent'' cuts is $n$. There is
one cut $[-x_0,x_0]$ centered around zero, and the other cuts come in pairs
$[x_{2i-1},x_{2i}]$ and $[-x_{2i},-x_{2i-1}]$. We shall follow these
notations in what follows.

We now demand that the resolvent $\omega_0(x)$ have the $S/x$ fall off
at infinity and thus get $n$ constraints 
\begin{equation}\label{falloffconstr}
\delta_{k,n}=
\half\cint{C}{x'}\frac{x'^{2k-1}W'(x')}{
\sqrt{\prod_{i=0}^{2(n-1)}(x'^2-x^2_i)}}
,\;k=1,2,\cdots,n.
\end{equation}
The most general solution to these $n$ constraints
(\ref{falloffconstr}) is given by  
\begin{equation}
g_n^2\prod_{i=0}^{2(n-1)}(x^2-x^2_i)=W'(x)^2-f(x),
\end{equation}
where $f(x)$ is the most general even polynomial of order $2n-2$
\begin{equation}
f(x)=\sum_{l=0}^{n-1}b_l x^{2l}.
\end{equation}

Note that we have now recovered the solution to the classical loop
equation that we obtained in section~\ref{sec:saddle}. We now repeat the
procedure outlined there and define the Riemann surface $\Sigma$ given by
\begin{equation}
y^2=W'(x)^2-f(x).
\end{equation}
The filling fractions $S_i$ then become period integrals of the
meromorphic 1-form $y\,dx$ over the 1-cycle $A_i$ of $\Sigma$ that
encircles the $i^\text{th}$ branch cut
\begin{equation}
S_i=\oint_{A_i}\frac{y\,dx}{2\pi i}.
\end{equation}
We can then argue that the change in the free energy due to bringing an
eigenvalue from infinity to the  $i^\text{th}$ cut is
\begin{equation}\label{bcycleint}
\pder{\CF_0}{S_i}=\int_{B_i}y\,dx.
\end{equation}
Note here that the $B$ cycles are non compact, so for (\ref{bcycleint})
to make sense we have to introduce an ultraviolet cut-off $\Lambda_{0}$
in the integral which has been identified with the bare coupling of the
gauge theory \cite{Cachazo:2002pr}. We comment here that there are only
semi-classical arguments for equation (\ref{bcycleint}), and we have been
unable to rigorously prove this as a consequence of the loop equations and
(\ref{dFbydV}).

\subsubsection{$\RP^2$ Contributions}
\label{sec:rptwocontr}

Once we have the form of the solution for $\omega_0(x)$, we can substitute
it in the loop equation, which is now a linear inhomogenous integral
equation for $\omega_1(x)$, 
\begin{equation}
\cint{C}{x'}\frac{W'(x')\omega_{1}(x')}{x-x'}
=\omega_{0}(x)\omega_{1}(x)-\frac{1}{2x}\omega_0(x).
\end{equation}

We can get a natural ansatz for $\omega_1$ from the string theory
expectation that $\CF_1$ should be a derivative with respect to $S_0$
of $\CF_0$, 
\begin{equation}\label{foneansatz}
\CF_1=q\pder{\CF_0}{S_0},
\end{equation}
where $q$ is some constant which has to be determined. Inserting this
into \eqref{wfromf}, we get 
\begin{equation}
\begin{split}
\omega_1(x)=&\oder{}{V}(x)\CF_1=
-q\sum_j\frac{2j}{x^{2j+1}}\pder{}{g_j}\pder{\CF_0}{S_0} \\
=&q\pder{}{S_0}\left(\omega_0(x)-\frac{S}{x}\right)\\
=&q\pder{\omega_0}{S_0}-\frac{2q}{x}.
\end{split}
\end{equation}

It is easy to see that $q\pder{\omega_0}{S_0}$ solves the homogeneous
part of the loop equation. The inhomogenous part of the loop equation
is solved by $-\frac{2q}{x}$ if $q=-\frac{1}{4}$.  

More generally, in the case of multi cut solutions, we could have
added any solution to the homogeneous loop equations. This amounts to
taking 
\begin{equation}
\CF_1=\sum_iq_i\pder{\CF_0}{S_i},
\end{equation}
such that $\sum q_i=-\frac{1}{4}$. However, corrections of the form
$\pder{\CF_0}{S_i}$ for $i>0$ should not be generated since these cuts
represent $U(N_i)$ gauge physics. We will give a short perturbative
discussion of this in the next section.  

We can extend this result to a single cut model with an arbitrary
polynomial potential. We will use this to solve the $SO/Sp$ $\CN=1^*$
theories.  

\subsection{Counting Feynman diagrams with $S^2$ and $\RP^2$ topology}

For a perturbative check of the relation
\begin{equation}
\CF_1 = \pm q \frac{\partial {\cal F}_0}{\partial S_0}
\label{g0}
\end{equation}
we need to enumerate ``ribbon'' graphs in the 't Hooft (genus)
expansion of the matrix model.  Recall that the genus expansion is
ordered by diagram topology, with diagrams of genus $g$ and $c$
cross-caps contributing at order $g_s^{-\chi} = g_s^{-2+2g+c}$.  The
coefficient $q$ is related to the relative contribution of the planar
(genus 0) diagrams which dominate at large $M$ with the leading
$\frac{1}{M}$ correction coming from diagrams with topology $\RP^2$.

It is known that $\SO(2M)$ and $\Sp(M)$ matrix models are related by
analytic continuation $M \mapsto -M$ (for the analogous gauge theory
results see \cite{Mkrtchian:1981bb,Cvitanovic:1982bq, Cicuta:1982fu}).
Therefore, at even orders in the 
genus expansion, the contribution to the matrix model free energy is
the same for both theories, while at odd orders the $\Sp(M)$ diagrams
contribute to the free energy with an additional minus sign relative
to $\SO(2M)$.  This fact determines the sign in \eqref{g0}.

\begin{figure}[tbp]
\begin{center}
\epsfig{file=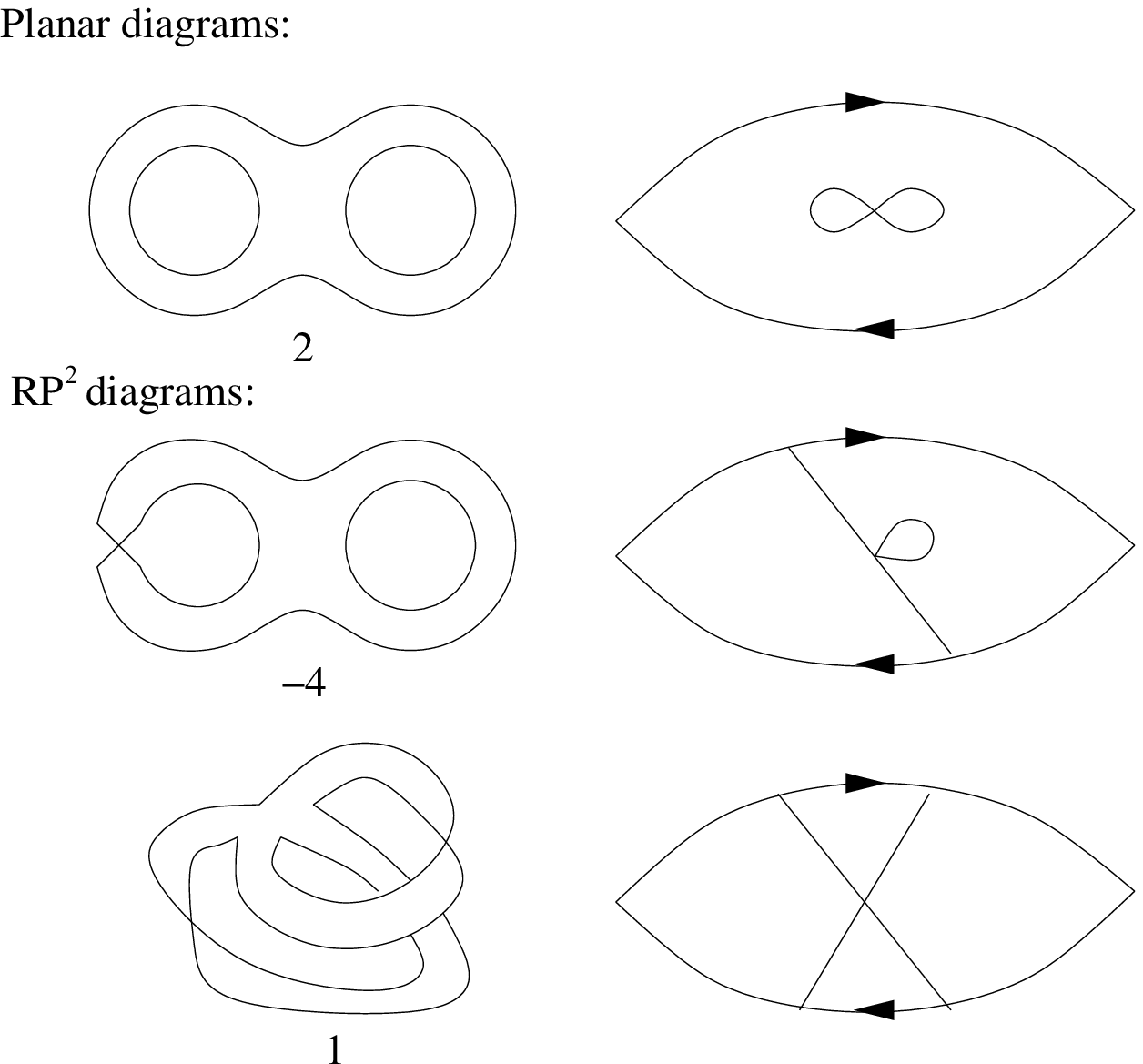}
\parbox{5.5in}{
\caption{$S^2$ and $\RP^2$ diagrams with one quartic vertex, written in terms of twisted and untwisted propagators and as diagrams on $\RP^2$ to show their
  planarity.  Propagators that pass through the cross-cap become
  twisted.\label{fig:diags}}\label{fig:diagrams}} 
\end{center}
\end{figure}

Recall that
\begin{equation}
\chi = v - p + l
\label{eq:chi}
\end{equation}
where $v$ is the number of vertices in the ribbon graph, $p$ is the
number of propagators and $l$ the number of boundary loops.  
The Feynman rules are summarized in appendix~\ref{app:feynman}.
Let us evaluate the first-order quartic diagrams in
fig. \ref{fig:diagrams}. The planar diagram has the value 
\begin{equation}
2\times\frac{1}{1!}\frac{g_2}{4g_s}\left(\frac{g_s}{2m}\right)^2M^3
\end{equation}
whereas the $\RP^2$ diagram with one twisted propagator contributes
\begin{equation}
-4\times\frac{1}{1!}\frac{g_2}{4g_s}\left(\frac{g_s}{2m}\right)^2M^2
\end{equation}
and the the $\RP^2$ diagram with two twisted propagators contributes
\begin{equation}
1\times\frac{1}{1!}\frac{g_2}{4g_s}\left(\frac{g_s}{2m}\right)^2M^2.
\end{equation}
Using the fact, that $S=\frac{g_s}{2}M$, this shows that
\begin{equation}
\CF_1=-\frac{1}{4}\pder{\CF_0}{S_0}
\label{eq:derrel}  
\end{equation}
at the first order. 

We have calculated the Feynman diagrams for several higher orders and
higher vertices and confirmed this relationship in those
cases\footnote{This relationship was apparently not known to
mathematicians.}. It would be nice to have a purely combinatorial proof
of~\eqref{eq:derrel},  that would  not rely on the loop equation. 

In order to describe a multi-cut matrix model, we need to use ghosts \cite{Dijkgraaf:2002pp}
to expand around the classical vacuum. In this prescription, one can
think of the matrix model as several matrix models, which are coupled
by bifundamental ghosts. Only one of those matrix models is actually
an $SO(M_0)/Sp(M_0/2)$ matrix model, the
other matrix models are $U(M_i)$ matrix models. The ghosts do not have
twisted propagators either, so the leading contribution from the
$SO(M_0)/Sp(M_0/2)$ matrix model is again the same as for a single cut
model. The loop equations still hold for the multi-cut model and we can extend the result
to all orders. 

\section{Computation of Effective Superpotentials}
\label{sec:superpot}

In this section we combine the results of the previous sections to
compute the effective superpotential of the dual gauge theories. We
will find that it is necessary to refine the formula for the
unoriented string contribution to the effective
superpotential of~\cite{Dijkgraaf:2002dh}.

\subsection{Non-Perturbative Sector}

As discussed in~\cite{Ooguri:2002gx,Dijkgraaf:2002dh}, there is a
non-perturbative 
contribution to the free energy which arises from the Gaussian integral:
\begin{equation}
{\cal F}^{np}
=\frac{1}{2} \text{dim}\, G \, \log\frac{2\pi g_s}{m}
-{\rm log\ vol}(G).
\end{equation}
In appendix B, following \cite{Ooguri:2002gx}, 
we have included the large $M$ expansion of the logarithm of the
volume of the $SO/Sp$ groups. We find that, for $SO(M)$ when $M$ is even, 
\begin{equation}
\begin{split}
{\cal F}^{np}=& \frac{1}{g_s^2} \CF^{np}_0 + \frac{1}{g_s} \CF^{np}_1
+ \cdots\\
= &  
\frac{1}{g_s^2} \left[ S^2 \log\frac{2\pi S}{m} 
- S^2 \left(\frac{3}{2} + \log \pi\right) \right] \\
& + \frac{1}{g_s} \left[ -\frac{S}{2} \log \frac{2\pi S}{m} 
+ \frac{S}{2} (1+\log\pi - \log 4 ) \right] + \cdots,
\end{split}
\end{equation}
with a similar expression for $M$ odd or $G=Sp(M)$. We see that 
\begin{equation}
{\cal F}_{1}^{np}= 
\mp \frac{1}{4}\frac{\del {\cal F}_{0}^{np}}{\del S} 
\pm \frac{1}{2} \log 2,
\end{equation} 
where the first $-/+$ sign is for $SO/Sp$ respectively. This is
almost the same relationship as we found for the perturbative
contributions~\eqref{eq:derrel}, but it is spoiled by the $\log 2$
term. We have traced this term through the volume computation outlined
in~\cite{Ooguri:2002gx} and found that it could be removed by a
different choice for the measure on the maximal torus of the Lie group. 

It is the non-perturbative sector, specifically the coefficient of the
$S^2\log S$ term, that determines the number of gauge theory vacua,
which is a main consistency test of the translation between matrix
model quantities and the effective superpotential of the gauge theory.
The number of vacua of a supersymmetric gauge theory is equal to the
dual Coxeter number $h$ of the gauge
group~\cite{Witten:1982df,Witten:1998bs}.   Therefore the total
superpotential should lead to the conclusion that $S^h$ is
single-valued.

Open string physics tells us that the sphere contribution to the
effective superpotential should be proportional to $Q_{D5}$, the total
charge of D5-branes, while the $\RP^2$ contribution should be
proportional to $Q_{O5}$, the total charge of O5-planes. We can
express this by refining the suggestion of~\cite{Dijkgraaf:2002dh}:
\begin{equation}
\Weff = Q_{D5}\, {\partial{\cal F}_0\over\partial S} + Q_{O5}\,\CG_0
-2\pi i \, \tau \, S,
\label{eq:WeffMM}
\end{equation}
We assume that  $\CG_0$ is proportional to the total $\RP^2$ free energy,
\begin{equation}
\CG_0 = a \left( \CF^{np}_1 + \CF^{p}_1 \right).
\end{equation}

Proceeding with this result, we find that 
\begin{equation}
\Weff = 
\left( \frac{N}{2} \pm \frac{a}{4} \right) S \log S - \frac{1}{2}
\tau\, S
+ \cdots,
\end{equation}
where the $+/-$ is for $SO/Sp$ respectively. 
Consistency with both the closed string result~\eqref{eq:weffclosed}
and the gauge theory\footnote{Note that, after including $a=\mp 4$, the
effective superpotential naively suggests that for gauge group
$Sp(N/2)$, $S^{N+2}$ is single-valued, whereas $h=N/2 +1$. The
resolution to this puzzle was explained in~\cite{Gomis:2001xw}. Namely
the D1-string wrapped on $\BP^1$ has instanton number {\it two}
in $Sp(N/2)$. Properly accounting for this reproduces the $\BZ_{2h}$
chiral symmetry of the dual gauge theory.}  requires that we must
have $a=\mp 4$. Very 
recently~\cite{Ita:Pert} produced this factor $|a|=4$ from a
perturbative argument along the lines
of~\cite{Dijkgraaf:2002xd}. It was found to be related to the measure
on the moduli space of Schwinger parameters, a quantity that is
intrinsic to the gauge theory. Presumably, there is a similar
explanation of this correction within the holomorphic Chern-Simons
theory. 

\subsection{The $\CN=1^*$ Theories}
\label{sec:onestar}

Following~\cite{Dijkgraaf:2002dh}, we can also consider the $\CN=1^*$
theories. Table~\ref {tab:dets} contains the results that are needed
to write the partition function  in the eigenvalue basis. In the large
$M$ limit, the discussion will entirely parallel that
of~\cite{Dijkgraaf:2002dh}. Inclusion of $\RP^2$ contributions to the
superpotential for $SO(2N)$ gauge group yields 
\begin{equation}
\Weff  =  \frac{2N-2}{2} \, \Pi_B(\tau) 
- \frac{1}{2}\tau_0 \, \Pi_A(\tau).
\end{equation} 
This has extrema at 
\begin{equation}
\tau = \frac{\tau_0 + k }{2N- 2}, ~k=0,\cdots  2N-1,
\end{equation}
at which points the superpotential takes the critical values
\begin{equation}
W_{(k)}  \sim E_2( (\tau_0 + k)/(2N-2)),
\end{equation}
in complete agreement with~\eqref{eq:soevenWcrit}.

\section{Discussion}
\label{sec:discussion}

In this paper we have outlined a general scheme for computing the
subleading contributions to the gauge theory effective superpotential
from $\RP^2$ diagrams in the dual matrix model of Dijkgraaf and
Vafa. The methods involve an application of the higher-genus loop
equations to determine the $\RP^2$ correction to the resolvent, which
allows us to compute the $\RP^2$ contribution to the free energy of
the matrix model. We then established a refinement of Dijkgraaf and
Vafa's relationship
between matrix model quantities and the gauge theory effective
superpotential which was necessary to obtain consistent field
theoretic results. The computation of~\cite{Ita:Pert} provides a gauge
theoretic explanation of our prescription.

There are many future directions that can be pursued. First, while we
were interested in an order $g_s$ (equivalently $1/M$) correction from
unoriented diagrams, in the matrix model duals to  gauge theory with
matter in general representations, such as quarks in the fundamental,
there will be $1/M$ corrections\footnote{For example, for $N_f<N_c$,
these correspond to terms of 
order $N_f/N_c$ in the gauge theory effective superpotential.} arising
from worldsheets with a single 
boundary. Our application of the loop equations should apply
to this case and it would be interesting to use this to make contact
with the results
of~\cite{Argurio:2002xv,McGreevy:2002yg,Suzuki:2002gp,Bena:2002kw,Demasure:2002sc}. 

Similarly, it would be of interest to examine higher-order corrections
in the genus expansion, which have an interpretation as gravitational
corrections~\cite{Dijkgraaf:2002fc,Dijkgraaf:2002dh,Klemm:2002pa,Dijkgraaf:2002yn}.
It would be interesting to make contact between the loop equations and
the Kodaira-Spencer equations
of~\cite{Bershadsky:1994cx}, which also relate higher genus results to
those at lower genus. It seems reasonable that there are higher genus forms
of our relation between oriented and unoriented contributions at a given genus like
\begin{equation}
\CF_{g,1}   \propto \pder{\CF_{g,0}}{S_0}. \label{eq:hgenusc}
\end{equation}
It would also be useful to obtain a deeper understanding of
relations like~\eqref{eq:hgenusc} from the diagrammatic
combinatorics. For example a diagrammatic proof of our conjecture seems to be possible.

\begin{acknowledgments}
The work of S.A. was supported in part by DOE grant number
DE-FG02-96ER40959
and that of R.C., N.H., K.K., and C.R.\ was supported in part by DOE
grant number DE-FG03-84ER-40168.
N.H.\ is supported by a Fletcher Jones Graduate Fellowship.
C.R.\ would like to thank the Aspen Center for Physics, the New High
Energy Theory Center at Rutgers, and the Department of Mathematical
Physics at the University of Adelaide for hospitality during the
course of this work.
We have greatly benefited from discussions with Bobby Acharya, Mina
Aganagic, Itzhak Bars, Per Berglund, Peter Bouwknegt, Andreas Brandhuber,
Michael Douglas, Pedro Fonseca, Jaume Gomis, Christiaan Hofman, Sameer
Murthy, Dennis Nemeschansky, Jongwon Park, Krzysztof Pilch, Alessandro
Tomasiello, and Nicholas Warner.
\end{acknowledgments}

\begin{appendix}

\section{Matrix Integral Measures and Determinants}

In this section we collect some results on the group measure and
adjoint action which are needed to do computations in the matrix models.

\subsection{The Group Measure for General Matrices}
\label{appvand}

We wish to compute the Jacobian for the transformation from
certain matrices $\Phi$ to their eigenvalues. This can be derived by a
group-theoretic
argument. In terms of the Cartan generators $H_i$ and ladder operators
$E_\alpha$, for the algebra of the group $G$, satisfying
\begin{equation}
[H^{i},E^{\alpha}] = \alpha^{i}E^{\alpha},
\end{equation}
we can diagonalize a matrix $\Phi$
\begin{equation}
\begin{split}
& \Phi = U^{\dagger}\Lambda U, \\
& \Lambda = \sum_{i}\lambda_{i}H^{i}.
\end{split} \label{eq:diagonalize}
\end{equation}
We will define parameters $t_\alpha$ so that
\begin{equation}
dU = \left[ \sum_\alpha\, dt_\alpha E_\alpha\right] U, 
~~~t_\alpha^*=-t_{-\alpha} .
\end{equation}

The infinitesimal variation of $\Phi$ can then be written as
\begin{equation}
\begin{split}
d\Phi = & U^{\dagger}
\left[d\Lambda + \sum_{\alpha}dt_{\alpha}\, [\Lambda,E^{\alpha}]\right]U
\\
= & U^{\dagger}
\left[d\Lambda + \sum_{\alpha}dt_{\alpha}\, 
\left( \sum_i\lambda_i\alpha^i\right)E^{\alpha}\right]U.
\end{split}
\end{equation}
We now calculate the metric on the Lie algebra
\begin{equation}\label{liemetric}
{\rm Tr}\, d\Phi \, d\Phi^\dagger
=  \sum_i d\lambda_i^2
+  \sum_{\alpha,\beta} dt_\alpha \,dt_\beta \,
\left(\sum_i\lambda_i \alpha_i \right)
\left(\sum_j\lambda_j \beta_j \right) \,
{\rm Tr}\,E_\alpha E_\beta.
\end{equation}
Using the identity
\begin{equation}
\textrm{Tr}_{r} E^{\alpha}E^{\beta} = C(r)\delta_{\alpha+\beta,0}
\end{equation}
where $C(r)$ is a representation dependent constant, we can simplify
the second term in equation \eqref{liemetric} to 
\begin{equation}
C(r)\sum_{\alpha} \left| \sum_i \alpha^i\lambda_i \right|^{2}
|dt_{\alpha}|^{2}
\end{equation}
Up to numerical factors, the Jacobian is
\begin{equation}
\Delta (\Lambda) = \prod_{\alpha > 0}
\left| \sum_{i}\alpha^i \lambda_i \right|^{2}.
\end{equation}

\subsection{The Induced Measure of the $\CN=1^*$ Theory}
\label{sec:detadj}

In this section, we calculate $\det\left(\adj_{\Phi} + im\right)$
whose modulus squared appears in the calculation of the induced
measure of the $\CN=1^*$ theory.  In order to calculate the
determinant, we go to a diagonal basis in which $\Phi$ is an element
of the Cartan subalgebra~\eqref{eq:diagonalize}. Then we solve the
eigenvalue equation
\begin{equation}
[\Phi,A] +i A = a A,
\end{equation}
where $A$ is a completely general matrix in the Lie algebra
\begin{equation}
A  = \sum_i h_i H^i 
+ \sum_\alpha t_\alpha E^\alpha. 
\end{equation}

We can compute
\begin{equation}
[\Phi,A] +i A = 
i  \sum_i h_i H^i 
+ \sum_\alpha \left( \sum_i \alpha^i\lambda_i +i\right) t_\alpha
E^\alpha, 
\end{equation}
so the eigenvectors and eigenvalues are
\begin{equation}
\begin{split}
&A = H^i , ~~a=i, \\
& A = E_\alpha ,~~ a = \bigl( \sum_i \alpha^i\lambda_i +i\bigr). 
\end{split}
\end{equation}
Up to numerical factors, the determinant is then 
\begin{equation}
\det\left(\adj_{\Phi} + im\right)\sim
\prod_{\alpha>0} \left(\sum_i \alpha^i\lambda_i+im\right).
\end{equation}

We list the expressions for the roots and the corresponding
determinants for the different classical groups in Table~\ref{tab:dets}. 

\medskip
\begin{table}[h]
\begin{tabular}{|l|c|}
\hline
$G$   & $J(\Lambda)$ \\
Roots & $\det(\adj_\Lambda+i)$ \\
\hline
$A_{N-1}$ & $\prod\limits_{i<j}(\lambda_i-\lambda_j)^2$ \\
$e_i-e_j$ ($i\ne j$) & 
$\prod\limits_{i<j}(\lambda_i-\lambda_j+i)(\lambda_i-\lambda_j-i)$ \\
\hline
$B_N$     & 
$\prod\limits_{i<j}(\lambda_i^2-\lambda_j^2)^2\prod\limits_i\lambda_i^2$ \\
$\pm e_i\pm e_j$ ($i\ne j$), $\pm e_i$ &
$\prod\limits_{i<j}((\lambda_i-\lambda_j)^2+1)((\lambda_i+\lambda_j)^2+1)\prod\limits_i(\lambda_i^2+1)$\\
\hline
$C_N$     & 
$\prod\limits_{i<j}(\lambda_i^2-\lambda_j^2)^2\prod\limits_i\lambda_i^2$ \\
$\frac{1}{\sqrt{2}}\left(\pm e_i\pm e_j\right)$ ($i\ne j$), $\pm \sqrt{2}e_i$ &
$\prod\limits_{i<j}((\lambda_i-\lambda_j)^2+2)((\lambda_i+\lambda_j)^2+2)\prod\limits_i(2\lambda_i^2+1)$\\
\hline
$D_N$     &
$\prod\limits_{i<j}(\lambda_i^2-\lambda_j^2)^2$ \\
$\pm e_i\pm e_j$ ($i\ne j$) &
$\prod\limits_{i<j}((\lambda_i-\lambda_j)^2+1)((\lambda_i+\lambda_j)^2+1)$\\
\hline
\end{tabular}
\caption{The roots and the formul\ae\ for the Jacobians and
determinants of the adjoint actions for the classical groups.}
\label{tab:dets}
\end{table}

\section{Asymptotic expansion of the gauge group volumes}

We now compute the asymptotic expansion of the volume of the gauge
groups, which normalizes the partition function of the matrix model
and provides the nonperturbative contribution to the free energy.  The
volumes are given by 
\cite{Ooguri:2002gx}:
\begin{equation}
\begin{split}
& \hbox{vol}(\SO(2N+1)) = {{2^{N+1}(2\pi)^{N^2 + N - {1 \over 4}}} \over {(2N-1)!(2N-3)!\ldots 3! 1!}}, \\
&\hbox{vol}(\SO(2N)) = {{\sqrt{2}(2\pi)^{N^2}} \over {(2N-3)!(2N-5)!\ldots 3! 1! (N-1)!}}, \\
&\hbox{vol}(\Sp(2N)) = {{2^{-N} (2\pi)^{N^2 + N}} \over {(2N-1)!(2N-3)!\ldots 3! 1!}}.
\end{split}
\end{equation}

We are interested in the large $N$ asymptotic expansion of the
logarithm of the volumes in order to compute the non-perturbative
contribution to the free energy.
Following \cite{Ooguri:2002gx}, we introduce the Barnes function
\begin{equation}
G_2(z+1) = \Gamma(z) G_2(z),\ G_2(1) = 1.
\label{barnes}
\end{equation}
Using the doubling formula for $\Gamma(z)$,
\begin{equation}
\Gamma(2z) = 2^{2z-1} \pi^{-{1 \over 2}} \Gamma(z) \Gamma(z + {1 \over 2}),
\label{gammadouble}
\end{equation}
and \eqref{barnes},  can evaluate the
denominator of the volume factors
\begin{equation}
G_d(N) \equiv (2N-1)!\ldots 3! 1! = {1 \over {(4 \pi)^{N/2}}} 2^{N(N+1)} G_2(N+1)G_2(N+{3 \over 2})
\end{equation}
Using the Binet integral formula
\begin{equation}
\log \Gamma(z) = (z - {1 \over 2}) \hbox{log} z - z + {1 \over 2} \hbox{ log\ }2 \pi + 2 \int_{0}^{\infty} {{\hbox{tan}({t \over z})} \over {e^{2 \pi t} - 1}} dt,
\label{binet}
\end{equation}
the asymptotic expansion of $G_2(n)$ is
\begin{equation}
\log G_2(N+1) = {N^2 \over 2} \log N - {1 \over 12} \log N - {3 \over 4} N^2 + {1 \over 2} N \log 2\pi + O(1).
\end{equation}
By expanding log($N-a$) for large $N$,  we obtain
\begin{equation}
\begin{split}
\log G_d(N) = & N^2 \log N + N^2(-{3 \over 2} + \log 2) \\
& + {1 \over 2} N \log N - {1 \over 24} \log N  + {N \over 2}(\log 4
\pi - 1) + O(1). 
\end{split}
\end{equation}

Putting all of this together, we find that
\begin{equation}
\begin{split}
& \log \hbox{vol}(\SO(2N+1))  \\
& \hskip2cm = -N^2 \log N+  N^2({3 \over 2} + \log \pi)  \\ 
& \hskip2.5cm - {1 \over 2} N \log N + {1 \over 24} \log N 
+ {N \over 2}(1 + \log 4 + \log \pi) + O(1), \\
& \log \hbox{vol}(\SO(2N)) \\
&\hskip2cm = -N^2 \log N + N^2({3 \over 2} + \log \pi) \\
& \hskip2.5cm+ {1 \over 2} N \log N + {1 \over 24} \log N 
+ {N \over 2}(-1 + \log 4 - \log \pi) + O(1), \\
& \log \hbox{vol}(\Sp(2N)) \\
& \hskip2cm = -N^2 \log N + N^2({3 \over 2} + \log \pi) \\
& \hskip2.5cm - {1 \over 2} N \log N + {1 \over 24} \log N 
+ {N \over 2}(1 - \log 4 + \log \pi) + O(1).
\end{split} \label{eq:asymptoticV}
\end{equation}

\section{Matrix model Feynman rules and enumeration of diagrams}
\label{app:feynman}

We want to perturbatively evaluate the matrix integral
\begin{equation}
\int d\Phi\, e^{\frac{1}{g_s}\Tr W(\Phi)},
\end{equation}
where the potential $W$ is given by
\begin{equation}
W(\Phi)=\sum_{j=1}^\infty\frac{g_j}{2j}\Phi^{2j}
\end{equation}
and $\Phi$ is a real antisymmetric $M \times M$ matrix.  We can write
this as
\begin{equation}
\int d\Phi \, \exp\left[
{\frac{1}{g_s}\Tr\left(\frac{m}{2}\Phi^2
+\sum_{j=2}^\infty\frac{g_j}{2j}\Phi^{2j}\right)} \right],
\end{equation}
where $m=g_1$.  Expanding the exponential leads to traces of integrals
of the form
\begin{equation}
\begin{split}
& \int d\Phi \, e^{\frac{1}{g_s}\Tr\frac{m}{2}\Phi^2}\, 
\Phi_{m_1n_1}\cdots\Phi_{m_kn_k}= \\
& \hskip2cm 
\pder{}{J_{m_1n_1}}\cdots\pder{}{J_{m_kn_k}}
\left(\int d\Phi \, \exp\left[{\frac{1}{g_s}\Tr\frac{m}{2}\Phi^2
-\half\Tr J\Phi}\right]\right)_{J=0}.
\end{split}
\end{equation}
This integral can now be evaluated, leading to
\begin{equation}
\left(\sqrt{\frac{2\pi g_s}{m}}\right)^{\frac{M(M-1)}{2}}
\pder{}{J_{m_1n_1}}\cdots\pder{}{J_{m_kn_k}}\left(e^{-\frac{g_s}{8m}\Tr J^2}\right)_{J=0}.
\end{equation}
Differentiating step by step gives rise to expressions like
\begin{equation}
\begin{split}
&\pder{}{J_{mn}}\left(\frac{g_s}{2m}J_{m_1n_1}\cdots\frac{g_s}{2m}J_{m_kn_k}e^{-\frac{g_s}{8m}\Tr J^2}\right) \\
&\hskip2cm
=\frac{g_s}{2m}(\delta_{mm_1}\delta_{nn_1}-\delta_{mn_1}\delta_{nm_1})\frac{g_s}{2m}J_{m_2n_2}\cdots\frac{g_s}{2m}J_{m_kn_k}e^{-\frac{g_s}{8m}\Tr
J^2} \\
&\hskip2.2cm+\cdots \\
&\hskip2.2cm+\frac{g_s}{2m}J_{m_1n_1}\cdots\frac{g_s}{2m}J_{m_{k-1}n_{k-1}}\frac{g_s}{2m}(\delta_{mm_k}\delta_{nn_k}-\delta_{mn_k}\delta_{nm_k})e^{-\frac{g_s}{8m}\Tr J^2} \\
&\hskip2.2cm+\frac{g_s}{2m}J_{mn}\frac{g_s}{2m}J_{m_1n_1}\cdots\frac{g_s}{2m}J_{m_kn_k}e^{-\frac{g_s}{8m}\Tr J^2}.
\end{split}
\end{equation}
The indices $m_i$ and $n_i$ are contracted in traces as given in the
interaction which can be interpreted as forming vertices.  The
combinatorics can the be interpreted diagrammatically, that one
must connect all the legs of the vertices in all possible ways with
untwisted and twisted propagators.  Each twisted propagator contributes
a factor of $(-1)$.

The rules for evaluating a diagram are then:
\begin{itemize}
\item Each kind of vertex with multiplicity $V_j$ contributes a factor
  of $\frac{1}{V_j!}(\frac{g_j}{2jg_s})^{V_j}$.
\item Each propagator contributes a factor of $\frac{g_s}{2m}$.
\item Each twisted propagator contributes a factor of $(-1)$.
\item Each index loop contributes a factor of $M=\frac{2S}{g_s}$.
\end{itemize}
The combinatorial factor of a diagram can be computed by counting all
topologically equivalent ways in which the legs of the vertices can be
connected.  This has some subtleties, since some diagrams with twisted
propagators can actually be planar.  To handle this, we make use of
the technique described in \cite{Cicuta:1982fu} to draw unoriented
diagrams (see also \cite{Mulase:2002cr,Mulase:2002xx} for recent work
on non-orientable ribbon diagrams in the mathematical literature).

An $\RP^2$ can be drawn in the plane as a disc, where antipodal
points on the boundary are identified.  $\RP^2$ diagrams can then be
drawn on that disc with some propagators going through the cross-cap
at the boundary.  The propagators going through the cross-cap are
twisted propagators, whereas all the others are untwisted propagators.

We can now also draw a planar diagram on the $\RP^2$.  If it has more
than one vertex, we can push one or several vertices through the
cross-cap without destroying the planarity, but all the propagators
going through the cross-cap are now twisted propagators.  This
operation contributes a multiplicative factor of $2^{v-1}$ to the
number of planar diagrams at each order $v$.  See Figure
\ref{fig:diags} for the enumeration of diagrams with 1 quartic vertex.

Using the relation between $p$ and the number of vertices $v_i$ of valency $i$ according to
\begin{equation}
p = \frac{1}{2} \sum_i i v_i
\label{eq:prop}
\end{equation}
the contribution of planar diagrams to the free energy of the $\SU(M)$
matrix model is given by
\begin{equation}
{\cal F}_{0} = \sum_{v=1}^{\infty} \frac{d^{(n)}_v}{v!} (\frac{g_n}{n g_s})^v (\frac{g_s}{m})^p M^l
= \sum_{v=1}^{\infty} \frac{d^{(n)}_v}{v!} (\frac{g_n}{n g_s})^v (\frac{g_s}{m})^{\half{n}v} M^{2-(1-\frac{n}{2})v},
\end{equation}
where the sum is over diagrams with $v$ vertices of valence $2n$,
$d^{(n)}_v$ is the number of planar diagrams at each order, and $l$ counts
the number of boundary loops of the ribbon graph.  The propagator for
$\SU(M)$ theories is twice that of the $\SO$/$\Sp$ theories.  In the
second line we have simplified using (\ref{eq:chi}) and
(\ref{eq:prop}).  The number of diagrams of topology $S^2$
(i.e.~planar diagrams) in $\SU(M)$ matrix theory with a quartic
potential is given by \cite{Brezin:1978sv}
\begin{equation}
d^{(4)}_{v} = \frac{(2v-1)! 12^v}{(v+2)!} 
= 2, 36, 1728, 145152, \ldots .
\end{equation}
We are not aware of explicit generating functions for other vertex
valences $2n$, but these diagrams can be enumerated by computer to the
desired order \cite{Kennaway:code}.

If we now include twisted propagators (i.e.~enumerate planar diagrams
in the $\SO$ or $\Sp$ matrix models), there is an extra contribution
to the set of planar diagrams coming from vertices that have been
``flipped'', converting untwisted to twisted propagators according to
the rule described above.

\begin{equation}
{\cal F}_{0} = \sum_{v=1}^{\infty} \frac{d^{(n)}_v}{v!} (\frac{g_n}{n g_s})^v (\frac{g_s}{2m})^p M^l
= \sum_{v=1}^{\infty} \frac{d^{(n)}_v}{v!} (\frac{g_n}{n g_s})^v (\frac{g_s}{2m})^{\half{n}v} M^{2-(1-\frac{n}{2})v},
\end{equation}
\begin{equation}
d^{(4)}_{v} = \frac{1}{2} \frac{(2v-1)! 24^v}{(v+2)!} 
= 2, 72, 6912, 1161216, \ldots .
\end{equation}
A similar expression exists for the $\RP^2$ free energy
\begin{equation}
{\cal F}_{1} = \sum_{v=1}^{\infty} \frac{\widetilde{d^{(n)}_v}}{v!} (\frac{g_n}{n g_s})^v (\frac{g_s}{2m})^p M^{l-1} 
= \sum_{v=1}^{\infty} \frac{\widetilde{d^{(n)}_v}}{v!} (\frac{g_n}{n g_s})^v (\frac{g_s}{2m})^{\half{n}v} M^{1-(1-\frac{n}{2})v} .
\end{equation}
Here the number of diagrams $\widetilde{d^{(n)}_v}$ is counted with a
minus sign for each twisted propagator\footnote{Gaussian Ensembles are
  matrix models that have been well-studied in the physics and
  mathematics literature.  The Gaussian Orthogonal and Gaussian
  Symplectic Ensembles also contain non-oriented ribbon diagrams with
  twisted propagators, however the propagator is $\langle T^a_b T^c_d
  \rangle \sim \delta_{ac} \delta_{bd} + \delta_{ad} \delta_{bc}$,
  {\it i.e.}, there is no relative minus sign between the two terms.  This
  corresponds to counting $\RP^2$ diagrams with a positive sign
  always.  Therefore the free energy of the Gaussian Ensembles differs
  from that of the Lie Algebra matrix models at sub-leading orders in
  the genus expansion.}.  The relevant planar and $\RP^2$ diagrams
were enumerated by computer up to 4 vertices with a quartic potential
$\Wtree \sim \Phi^4$, to 2 vertices with a sextic potential $\Wtree
\sim \Phi^6$, and for a single vertex with a potential of degree up to
16.  The results are summarized in Table \ref{tab:enum} and verify the
desired relation:

\begin{equation}
{\cal F}_1 = -\frac{1}{2} \frac{\partial {\cal F}_{0}}{\partial M}.
\end{equation}

\begin{table}[h]
Diagrams with quartic vertices:
\begin{center}
\begin{tabular}{|c|c|r|r|r|r|}
\hline
Gauge group & Topology & $v=1$ & $v=2$ & $v=3$ & $v=4$ \\ \hline \hline
$\SU$ & $S^2$ & $2 M^3$ & $36 M^4$ & $1728 M^5$ & $145152 M^6$ \\ \hline
$\SO$/$\Sp$ & $S^2$ & $2 M^3$ & $72 M^4$ & $6912 M^5$ & $1161216 M^6$ \\
$\SO$/$\Sp$ & $\RP^2$ & $-3 M^2$ & $-144 M^3$ & $-17280 M^4$ & $-3483648 M^5$ \\
\hline
\end{tabular}
\end{center}

Diagrams with sextic vertices:
\begin{center}
\begin{tabular}{|c|c|r|r|r|}
\hline
Gauge group & Topology & $v=1$ & $v=2$ \\ \hline \hline
$\SU$ & $S^2$ & $5 M^4$ & $600 M^5$ \\ \hline
$\SO$/$\Sp$ & $S^2$ & $5 M^4$ & $1200 M^6$ \\
$\SO$/$\Sp$ & $\RP^2$ & $-10 M^3$ & $-3600 M^5$ \\ \hline
\end{tabular}
\parbox{5.5in}{
\caption{\small{Contribution to the free energy of the $\SU$/$\SO$/$\Sp$ matrix models at planar and $\RP^2$ level, for quartic and sextic potentials.  The first few terms in the perturbative expansion are listed, corresponding to the number of diagrams with increasing number of vertices (equivalently loops).}}
\label{tab:enum}}
\end{center}
\end{table}

\end{appendix}

\vfill
\providecommand{\href}[2]{#2}\begingroup\raggedright\endgroup

\end{document}